\documentclass{article}
\usepackage{graphicx}
\usepackage{amsmath}
\usepackage{amsfonts}
\usepackage{amssymb}
\begin{document}
\begin{center}
{\bf\large The Numerical Solution of Nekrasov's Equation in the Boundary
Layer near the Crest, for Waves near the Maximum Height}

\end{center}
\medskip
\begin{center}
{\bf by J.G. Byatt-Smith}

\end{center}
\medskip

\begin{center}
{\bf Department of Mathematics and Statistics}

{\bf University of Edinburgh.}\bigskip 
\end{center}


\noindent \textbf{Abstract}:\ \ Nekrasov's integral equation describing
water waves of permanent form, determines the angle $\phi\left( s\right)
$ that the wave surface makes with the horizontal.  The independent
variable $s$ is a suitably scaled velocity potential, evaluated at the free
surface, with the origin corresponding to the crest of the wave. For all
waves, except for amplitudes near the maximum, $\phi\left( s\right) $
satisfies the inequality $|\phi\left( s\right) |<\pi/6$. 

It has been shown numerically and analytically, that as the wave amplitude
approaches its maximum, the maximum of $|\phi\left(  s\right)  |$ can exceed
$\pi/6$ by about $1\%$ near the crest. Numerical evidence suggested that
this occurs in a
small boundary layer near the crest where $\left|  \phi(s)\right|$ rises
rapidly from $|\phi\left(  0\right)  |=0$ \ and oscillates about $\pi/6$,
the number of oscillations increasing as the maximum amplitude is approached.

 McLeod derived, from Nekrasov's equation, the following integral equation
\[
\phi\left(  s\right)  =\frac{1}{3\pi}\overset{\infty}{\underset{0}{\int}
}\frac{\sin \phi \left(  t\right)  }{1+\int_{0}^{t}\sin \phi \left(  \tau\right)  d\tau
}\log\left |\frac{s-t}{s+t}\right |dt
\]
 for $\phi\left(  s\right)  $ in the boundary layer, whose width tends to zero as the
maximum wave is approached. He also conjectured that the asymptotic form of
$\phi\left(  s\right)  $ as $s\rightarrow\infty$ satisfies
\[
\phi\left(  s\right)  =\frac{\pi}{6}\left\{ 1 +As^{-1}\sin\left(\beta \log 
s+c\right)+ o(s^{-1})\right\}, 
\]
where $A,\beta$ and $c$ are constants with $\beta $ $\approx $ $0\cdot71$ 
 satisfying  the equation
\[
\sqrt{3}\beta\;\tanh\;\frac{1}{2}\pi \beta=1.
\]
We solve McLeod's boundary layer equation numerically and verify the
above asymptotic form.
\newpage
\section{Introduction\strut}

This paper considers the numerical solution of the equation
\begin{align}
\phi\left(  s\right)   &  =\frac{1}{3\pi}\underset{0}{\overset{\infty}{\int}
}\frac{\sin\phi\left(  t\right)  }{1+\int_{0}^{t}\sin\phi\left(  \tau\right)
d\tau}\log\left|  \frac{s+t}{s-t}\right|  dt\tag{1.1a}\\
&  =-\frac{1}{3\pi}\underset{0}{\overset{\infty}{\int}}k\left( 
t,s\right)  
\left\{  \psi\left(  t\right)  -\psi\left(  s\right)  \right\}  dt\text{,}
\tag{1.1b}
\end{align}

where
\begin{equation}
\psi\left(  t\right)  =\log\left(  1+\int_{0}^{t}\sin\phi\left(  \tau\right)
d\tau\right) \mbox{ and } k\left(  t,s\right)  =\frac{2s }{s^{2}-t^{2}} 
.  \tag{1.2}
\end{equation}
 This equation was derived by McLeod [1] to describe the boundary
layer behavior of the solution , for large $\mu$, near the origin of the equation
\begin{equation}
\phi_{\mu}\left(  s\right)  =\frac{1}{3\pi}\underset{0}{\int^{\pi}}\frac
{\sin\phi_{\mu}\left(  t\right)  }{\mu^{-1}+\int_{0}^{t}\sin\phi_{\mu}\left( 
\tau\right)
d\tau}\log\left|  \frac{F\left(  s+t\right)  }{F\left(  s-t\right)  }\right|
dt\text{,} \tag{1.3}
\end{equation}
 where $F\left(  t\right)  =\mbox{sn}\left(  Kt/\pi\right)  $ and
$\mbox{sn}$ denotes the Jacobian elliptic function with quarter periods $K$ and
$iK^{\prime }$. Equation (1.3) was first formulated by Nekrasov [2] to describe
waves of constant periodic form moving with constant speed on the surface of a
non-viscous fluid that is either of infinite depth or on a horizontal 
bottom, when the flow is taken to be irrotational. The wave is assumed
to be symmetric about its crest and the equation is
derived by conformally mapping the the region of the flow under one
wavelength  onto the unit disc cut along the negative real axis. The generic
point on the circumference of the disc is $e^{is}$, with $-\pi < s <\pi
$, and  $s=0$ corresponds to the crest. As the circumference is
described in a clockwise direction from $-\pi
$ to $\pi $ the horizontal coordinate decreases by one wavelength. Then
the function  $\phi_{\mu}$ is the angle that the wave surface makes with
 the horizontal. With this choice of  coordinate $\phi_{\mu}(s)$  is periodic
with period $2\pi$.  For more details, see Nekrasov [2],[3] and [4] or
Milne-Thompson [5].   
The wave is assumed to be symmetric about its crest. 
Thus $\phi_{\mu}\left(  s\right)  $ is
an odd $2\pi$ periodic function of $s$ with $\phi_{\mu}\left(  0\right)  =0$.
The solution is unique provided the additional assumption, that the wave has
only one peak and one trough per period, is made. This is 
\begin{equation}
\phi_{\mu}\left(  s\right)  >0,\;s\in\left(  0,\pi\right)  \;\text{with }
\phi_{\mu}\left(  0\right)  =\phi_{\mu}\left(  \pi\right)  =0. \tag{1.4}
\end{equation}
 The constants $K$ and $iK^{\prime }$, the quarter periods of sn, are
related to the depth $h$ and wavelength, $\lambda$, by the relation
\begin{equation}
K^{\prime }/K=h/\lambda. \tag{1.5}
\end{equation}
 As $h\rightarrow\infty$ we have $K\rightarrow\frac{1}{2}
\pi\;\left(  K^{\prime }\rightarrow\infty\right)  $ and $F\left(  t\right)
\rightarrow\;\sin\;\frac{1}{2}t$ so that (1.3) is also applicable for infinite
depth. Equation (1.1) is derived by writing $\hat{s}=s\mu$ and writing
$\phi_{\mu}\left(  s\mu\right)  =\hat{\phi}\left(  \hat{s}\right)  $ and
letting $\mu\rightarrow\infty$ with $\hat{s}$ fixed. Then $\hat{\phi}\left(
s\right)  $ satisfies (1.1). The boundary layer behavior of the solution of
(1.3) was established numerically by Chandler and Graham [6], who were able
to obtain a solution with a maximum value of $\phi_{\mu}\left(  s\right)
\bumpeq30\cdot3787\ldots^\circ $ and to detect a small number of oscillations
about $\phi_{\mu}=30^\circ $ for $\mu=10^{18}$.

The numerical difficulty posed by the boundary layer behavior of the solutions
of (1.3) for large $\mu$ is over come, by Chandler and Graham [6], by using a non
uniform mesh for the discretisation of (1.3). This consists of three regions:
one to cope with the rapid variation of $\phi_{\mu}\left(  s\right)  $ in the
boundary layer, whose thickness is of order $\mu^{-1},$ near the origin; a
second to deal with the slower variation away from the origin and a third for
the transitional layer in between. For further references on the analytical
properties of the solutions of (1.3) and related numerical results, see
Chandler and Graham [6] and McLeod [1].

The purpose of this paper is to solve (1.1) numerically and show that the
solution $\phi\left(  s\right)  $ oscillates about $\phi\left(  s\right)
=\pi/6$ and obeys the formal asymptotic result of McLeod [1] that can be
written in the form 
\begin{equation}
\phi\left(  s\right)  =\frac{\pi}{6}\left\{ 
1+\overset{\infty }{\underset{n=0}{\sum}} \frac{A_n}{s^n}\sin\left(  n\beta\log
s+c_n\right)\right\}  \text{ as }s\rightarrow\infty\text{,} \tag{1.6}
\end{equation}

where $A_n$ and $C_n$ are constants and $\beta=0\cdot 71 \ldots $ is the root of
\begin{equation}
\sqrt{3}\beta\tanh\left(  \frac{1}{2}\pi\beta\right)  =1\text{.} \tag{1.7}
\end{equation}

Equation (1.1) represents the solution in the boundary layer and can thus be
solved with a uniform mesh size. However (1.1) has an additional complication
compared with (1.3) in that the range of integration is infinite and the decay
of the solution to its asymptotic limit is algebraic. This fact means that we
require careful consideration in order to obtain an accurate numerical
representation of the integral in (1.1).

\newpage
\section{The Numerical Method}

Following Chandler and Graham [6] we solve the integral equation in the form
(1.1b). This formulation is better, for numerical purposes, because the
integration by parts that is used to convert (1.1a) to (1.1b), removes the
logarithmic singularity, at $t=s,$ which occurs in the kernel of (1.1a).
Although the corresponding kernel of (1.1b) has a pole, the singularity of the
integrand is removable since the multiple $\psi\left(  t\right)  -\psi\left(
s\right)  $, has a simple zero at $t=s.$

Thus we write
\begin{equation}
\phi\left( s\right) =\frac{1}{3\pi}\int_{0}^{\infty}K\left( t,s\right)
dt, \tag{2.1}
\end{equation}
 where
\begin{align}
K\left(  t,s\right)   &  =-\frac{2s\left(  \psi\left(  t\right)  -\psi\left(
s\right)  \right)  }{s^{2}-t^{2}} & t\neq
s \ \tag{2.2a}\\
&  =\psi^{\prime }\left(  t\right)  \equiv \frac{\sin \phi \left(  t\right)
 }{1+\int_{0}^{t}\sin \phi\left(  \tau\right)  d\tau
} & t=s, \tag{2.2b}
\end{align}
 the value in (2.2b) being the limit of the right hand side
of (2.2a) as $\left|  t-s\right|  \rightarrow0$.

We aim to set up a numerical approximation to the integral in terms of a
discrete number of values $\phi\left(  s_{i}\right)  $, where $s_{i}
=ih$, $0\leq i\leq2N$, with $N$  an integer, for suitable choices of $h$
and $N$ and a continuous set of values $\phi\left(  s\right)  $ for
$s\geqslant2Nh$. Any values of $\phi\left(  s\right)  $ for $s<0$ required by
the numerical approximation are determined by the fact that $\phi\left(
s\right)  $ is an odd function of $s$.
The numerical representation of the integral requires two approaches. The
first is a finite difference formulation of the integral over a predetermined
finite range using the discrete values of $\phi$ and the second is an
estimation of the remainder using an appropriate asymptotic estimate of the
values of $\phi\left(  s\right)  $ for $s\geqslant2Nh$. The details of the
asymptotic form of $\phi\left(  s\right)  $ as $s\rightarrow\infty$ that is
used will be discussed later.

So we choose an appropriate end point $2T$ where $T$ is given by $T=Nh$
and we can approximate the integral $I_{1}\left( s,\phi\right)
=\int_{0}^{2T}K\left( t,s\right) dt$ using Simpson's Rule, since the
integrand is analytic.  The choice of the end point $2T$ is some what
arbitrary.  Eventually, see below, we will want to consider $I_{1}\left(
s,\phi\right) $ for values of $s\leq T$.  We choose an end point $mT$,
with $m=2$ in this case, so that the singularity of $k\left( t,s\right)
$ at $t=s$ is far from the end point.  The reason for this is that the
remainder integral, again see below, requires a different evaluation and
it is advantageous to make sure that the singularity of $k\left(
s,t\right) $ is not close to the range of $t$ in the remainder integral. 
This will become clearer when the evaluation of the remainder integral
is discussed later. 

Assuming that for large $s,\ \phi\left(  s\right)  $ is known in the form of an asymptotic
expansion then truncation of this series, expansion of the integrand and a
term by term integration of the integrand will give a suitable analytical
estimate $EI_{2}\left(  s,\phi\right)  $ for the integral $I_{2}\left(
s,\phi\right)  =\int_{2T}^{\infty}K\left(  t,s\right)  dt$. Then we define the
numerical representation of the integral in (2.1)
as 
\begin{equation}
NI\left(  s,\phi\right)  =NI_1\left(  s,\phi\right)  +EI_{2}\left(
s,\phi\right)  . \tag{2.3}
\end{equation}
 An alternative approach, assuming that the asymptotic form of
$\phi\left(  s\right)  $, $s>T$, has been chosen, is to transform the
infinite range of the remainder integral into a finite range, which can then
be approximated numerically. For this purpose it is more convenient to revert
to the integral in the form (1.1a) so we write
\begin{equation}
I_{2}\left( s\right) =\log\left( \frac{2T+s}{2T-s}\right) \left(
\psi\left( 2T\right) -\psi\left( s\right) \right)
+\overset{\infty}{\underset{2T}{\int}}k_{3}\left(   
s,t\right)  dt, \tag{2.4}% 
\end{equation}
 where 
\begin{equation}
k_{3}\left( s\right)= \frac{\sin\phi\left( t\right)
}{1+\int_{0}^{t}\sin\phi\left( \tau\right) d\tau}\log\left(
\frac{t+s}{t-s}\right) \tag{2.5}
\end{equation}
 If $\ \phi\left( t\right) \to \ \pi /6 +O(t^{-1})$ and
$\int_{0}^{\infty} (\phi\left( t\right)-\pi /6) dt$ is bounded, it is
easily established that  $k_{3}\left( s,t\right) =2st^{-2}+o\left(
t^{-2}\right) \;\text{as }t\rightarrow \infty .$
 Thus the integral of $k_{3}$, in (2.4) is convergent at infinity and
the substitution  $t=2T/u$ transforms it to
${\int}_{0}^{1}k_{4}\left( s,u\right) du $
 with $k_{4}\left(  s,0\right)  =s/T.$
This integral can  now be approximated using Simpson's rule 
with a suitably chosen step length. This approximation can be used
instead of $EI_{2}\left(  s,\phi\right)  $ in (2.3).

Simpson's rule gives an approximation which is of order $h^{4}$, but
this rule requires an interval which consists of an even number of step
lengths.  However the integrand contains the function $\psi\left(
t\right) $ which involves the determination of
$\int_{0}^{t}\sin\phi\left( \tau\right) d\tau$ at values $t=t_{i}=ih$. 
To obtain a numerical approximation to this which is the same order as
Simpson's rule for this integral we use an appropriate modified
trapisoidal rule.

 We now wish to solve the approximation
\begin{equation}
\phi\left(  s\right)  =\frac{1}{3\pi}NI\left(  s,\phi\right).  \tag{2.6}
\end{equation}
 To do this we define an approximation $\phi_{N}\left(  s_i\right)  $
to the solution $\phi\left(  s\right)  $ at the discrete values $s_{i}
=ih,\;0\leq i\leq N$. Using the same asymptotic form at the solution as that
used to define $\phi\left(  s\right)  $ for $s\geqslant2Nh$ we define the
remaining discrete values of $\phi_{N}\left(  s_i\right)  $, $N+1\leq i<2Nh$,
required for the evaluation of $NI_{1}\left(  s\right)  $ at the points
$s=s_{i}$, $0\leq i\leq N$.

Thus $\phi_{N}\left(  s_{i}\right)  $ satisfies the equations
\begin{equation}
\phi_{N}\left(  s_{i}\right)  =\frac{1}{3\pi}NI\left(  s_{i}
,\phi_{N}\left(  s_j\right)  \right)  ,\;0\leq i\leq N.\tag{2.7}
\end{equation}
 This gives, in a similar fashion to Chandler and Graham [6], a fully
discrete non-linear system for the unknowns $\left\{  \phi_{N}\left(
s_i\right)  ,i=0 ..N\right\}  $. This system is solved by the iterative
method 
\begin{equation}
\phi_{N}^{m}\left(  s_i\right)  =NI\left(  s_{i},\phi_{N}^{\left(  m-1\right)
}\left(  s_{j}\right)  \right)  ,\;i=0..N, \tag{2.8}
\end{equation}
 starting from a suitable initial approximation $\phi_{N}^{\left(
0\right)  }\left(  s_{i}\right)  $. Chandler and Graham [6] were able to prove
that, when the quadrature method used to approximate their integrals was the
trapisoidal rule, convergence was guaranteed, although for computational
purposes they opted for a more accurate scheme for computational purposes.
Their proof cannot be extended to the numerical approximation used here even
if the quadrature method is the trapisoidal rule because of the infinite range
of integration. However we find that, as in the cases looked at by Chandler
and Graham [6], the convergence rule is very quick.

\newpage
\section{The necessity of rescaling}

We see from the definition of $K\left( t,s\right), \;$(2.2a,b), and the
fact that $\phi\left( 0\right) $ is zero, that $NI\left( 0,\phi\right)
=0$ provided the initial guess $\phi_{N}^{\left( 0\right) }\left(
0\right) =0$.  Then (2.7) gives $\phi_{N}^{m}\left( 0\right) =0$ for all
$m>0$.  Thus effectively we can work with the $N$ variables $\left\{
\phi_{N}\left( s_i\right) ,i=1..N\right\} $ and corresponding $N$
equations from (2.7).  One of the aims is to verify the asymptotic
result (1.6).  Initially we do not assume this and report here that for
a variety of sensible choices of the asymptotic form of $\phi\left(
s\right) $ we get rapid convergence to the solution of (2.8).  Provided
$T$ is sufficiently large we can then numerically verify that (1.6) is
the correct asymptotic result, using the computed values of $\phi\left(
s\right) $ for $s\leq T$.  Having verified this numerically to get the
best accuracy we use (1.6) and  find that as well as providing a more
accurate numerical solution the convergence rate is also improved.  
The larger $T$ is, the less necessary it is to have a large
number of terms from (1.6) and in practice we use 
\begin{equation}
\phi\left(  s\right)  =\frac{\pi}{6}\left(  1+\frac{A}{s}\sin\left(  \beta\log
s+c\right)  \right)  ,\;\;s>T. \tag{3.1}
\end{equation}
 Table 1 shows the comparison of the location and the values of
$\phi\left(  s\right)  $ at successive maximum and minimum values of $\phi$
and the comparison between this method at that of Chandler and Graham [6]. Before
discussing this comparison we use the values of $s$ at the successive turning
points to illustrate the need for rescaling the variable $s$. It will become
clear that the computations done to obtain table 1 could not be achieved by
the method outlined in paragraph 1. We see that the s coordinate of each
successive turning
point increases by a factor of about $81$, which is approximately the
value of $e^{\pi/\beta}$ . This is compatible with the set of turning
points obtained from (3.1).
 The last turning point in $0<s<T$ is located at $s=2\times10^{11}$. Typically
we used $h=1/20$ as a sensible choice of $h$ compatible with having a large
enough $T$ to capture the asymptotic behavior of the solutions. However with
this choice of $h$ it is not feasible to take $T=2\times10^{11}$ as this would
involve $4\times10^{12}$ grid points. Typically  using the scheme outlined 
in paragraph 1 we chose $T=100$ and this does not even get to the first minimum
of $\phi\left(  s\right)  $. However we learn from this initial attempt at a
numerical solution that beyond $s=100,$ $6\left|  \phi\left(  s\right)
-\pi/6\right|  /\pi<10^{-2}$ and varies very slowly. Thus for large $s$ we do
not need to take such a small step length.

For the numerical scheme we have used, we require a constant steplength
so we make a simple change of independent variable. We wish to make no
effective change at the origin but an exponential change at infinity so we
use the transformation $s=e^{y}-1.$
 Then with $t=e^{z}-1$ and $\theta\left(  y\right)  =\phi\left(
s\left(  y\right)  \right),  $ (1.1) becomes 
\begin{equation}
\theta\left(  y\right)=-\frac{1}{3\pi}\int_{0}^{\infty}\frac{2\left(  e^{y}-1\right)  }{\left(
e^{y}-e^{z}\right)  \left(  e^{y}+e^{z}-2\right)  }\log\left(\dfrac
{1+\int_{0}^{z}\sin\theta\left( 
\zeta\right)}{1+\int_{0}^{y}\sin\theta\left(  \zeta\right)}\right )   dy, \tag{3.3}
\end{equation}
 We are then able to reduce the step length, $h$, and still take
$T=e^{y_T}-1$ to be large. Typically we take $h=1/100$ and $y_{T}=30$ giving
$T=1.0\times10^{13}$. This requires $3000$ unknowns $\phi\left(  y_i\right)  $
where $y_i=ih,\;\;i=1\;..\;3000$.

After the rescaling, the numerical scheme is essentially the same as that given
in section 2 and is not repeated. However near $y=y_{T}$, $6\left|  \theta\left(
y\right)  -\pi/6\right|  /\pi$ is now of order $10^{-13}$ so the form of
$\theta\left(  y\right)  $ effectively given by (3.1) will be accurate to
$10^{-26}$, that is $O\left(  T^{-2}\right)  $.

\newpage
\section{The Numerical Results and Conclusions}

All the numerical results given here are those produced by the numerical
scheme outlined in Section 2 and 3 using the rescaled problem. Table 1 shows
the comparison of the successive maxima and minima of $\phi\left(  s\right)  $
compared with those computed for the full problem by Chandler and Graham [6]. The
position of these maxima and minima for the Chandler and Graham [6] computation,
has been calculated by scaling their coordinate, $s$, by $\mu$ compatible with
the boundary layer scaling used to derive (1.1) from (1.3). Thus $s=s_{B-S}
=s_{CeG}\times\mu$. The number of decimal places given in table 1 for this numerical
computation are as accurate as the numerical calculation will allow. There are
three forms of error: the first comes from the order of the numerical
approximation to the solution which is $O\left(  h^{4}\right)  $ which gives
rise to errors of order $10^{-8}$; the second is due to machine accuracy which
gives rise to an error of about $10^{-14}$ to $10^{-16}$; thirdly there is the
error that arises when predicting the position and size of the maxima and
minima of a function, from discrete data at given grid points, assuming that
the data is accurate. The figures quoted in table 1 do not take into account
the first of two of these sources of error.

The comparison with the computations of Chandler and Graham [6] is very good. The
value at the first maximum is the same to eight significant figures and the
position the same to six significant figures. The calculation of the value at
the maximum always being more accurate that its positions. The values at the
first minimum are in similar agreement although Chandler and Graham [6] only quote
the position to four significant figures and the value at the minimum is only
$4\times10^{-3}$ below $30^{\circ}$ so relatively the numbers do not appear to
be in such good agreement as the value at the first maximum. The first
noticeable divergence of the two computations appears at the second minimum
where the estimates of the positions differ by about $4\%$ although the values
at this minimum are in good agreement given that they are both of order
$10^{-7}$ below $30^{\circ}.$ However the next maximum of Chandler and Graham [6]
lies below $30^{\circ}$ and it is apparent that at this value of $s$ the
effects of the outer solution, that is the  decrease from the maximum on
a slower scale, are just beginning to show. Presumably at this
value of $\mu$ the oscillations in the Chandler and Graham [6] begin to cease at
or around this value of $s$.

We wish to show that the solution behaves like (1.6) for large $s$. So for
comparison we write $\Theta\left(  x\right)  =\phi\left(  s\right)  $, where
$x=\frac{\beta}{\pi}\log s  $ so that we expect 
\begin{equation}
 \Theta\left(  x\right)  \sim \frac{\pi}{6}\left\{  1+\frac{A}{s}
\sin\pi\left(  x-x_{0}\right)  +\ldots \right\} 
\text{as }x    \rightarrow+\infty \tag{4.1}
\end{equation}
 or 
\begin{equation}
\Psi\left(  x\right) \equiv \left(  \frac{6}{\pi}\Theta\left(  x\right)  -1\right)
s \sim A\sin\pi\left(  x-x_{0}\right) + \ldots,  \tag{4.2}
\end{equation}
 Compared with the transformation (3.1) which has $y=0$ when $s=0$
we have $x\rightarrow-\infty$ as $s\rightarrow0$. This makes $\left(  \frac
{6}{\pi}\Theta\left(  x\right)  -1\right)  s\rightarrow0$ as $x\rightarrow
-\infty$ and introduces a minimum of the function $\Psi\left(  x\right)  $
before the first maximum. The values of $x=x_{i}$ at the minima, maxima and
the zeros of $\Psi\left(  x\right)  $ and the value of $\Psi\left(  x\right)
$ at the turning points are shown in table 2. If (4.2) were to be exact then the
difference $x_{i}-x_{i-1}-1/2\equiv\Delta x_{i}$ would be zero and the
magnitude of the value of $\Psi\left(  x\right)  $ at the turning points would
be constant and equal to $A$. Included in this table are the computed values
of $\Delta x_i.$

From the table we see that a good fit is obtained by choosing $A$ and $x_{0}$
so that $\Psi\left(  x\right)  $ and (4.2) agree at the second maximum and
fourth zero this gives
\begin{equation}
A=1\cdot2364860386 \ldots \mbox{ and } \tau_{0}=0\cdot72422 \ldots \ .\tag{4.3}
\end{equation}
 A plot of the asymptotic expression (4.2) with these values of $A$ and
$x_{0}$ and the comparison with $\Psi\left(  x\right)  $ is given in figure 1.
The two graphs are indistinguishable from each other over a surprisingly large
range of values of $x$, from before the first zero to beyond the sixth zero.
The graphs start to diverge after this point. This is due to the fact that the
exact solution of $\phi\left(  s\right)-\pi/6,  $ or equivalently
$\Psi(x)/s, $ is so small in this range that
round off error starts to become important and eventually dominates the
solution. This is more apparent in figures 2 and 3 which plot the difference
between $\Psi\left(  x\right)  $ and its asymptotic value. Figure 2 shows this
difference multiplied by $100$ in the range of values of $x$ where the
difference is less
then one, while figure 3 shows $1000$ times the difference. In both figures we
see that the difference increases rapidly after $x\bumpeq4$. It is
particularly visible in figure 3 that this rapid rise has two different
components: a systematic rise due to truncation error of the numerical scheme,
which is of order $10^8$ and a random error on the scale of about $10^{-14}$,
due to machine accuracy.

The last plot, figure 4, shows the difference between $\Psi\left(
x\right) $ and its asymptotic value multiplied by $s$.  This clearly
shows that the dominant feature is one of a periodic function of period
1, compatible with a term proportional to $s^{-2}\sin2\pi\left(
x-x_{1}\right) $ that appears in (1.6). 

To conclude we have presented a numerical scheme for the solution of (1.1),
written in the form (3.3) which allows a sufficiently accurate numerical solution
over a range $0\leqslant s\leqslant10^{13}$, that we can verify the predicted
asymptotic form (1.6). The numerical calculation is limited by the two factors,
truncation error and machine accuracy. The numerical solutions can be made
more accurate by a higher order integration scheme but the range of
integration is limited because the difference between the solution and $\pi/6$
becomes the same order of magnitude as the machine accuracy.

\newpage

\noindent \textbf{References}

\noindent 1. J.B. McLeod,
The Stokes and Krasovskii Conjectures for the wave of greatest height.
\textit{Stud. App. Math.} 98: 311-333 (1997)\medskip

\noindent 2. A.I. Nekrasov, \textit{Izv. Ivanovo-Vosnosonk.
Politehn Inst.} 3: 52-65 1921; 6:155-71 (1922)\medskip

\noindent 3. A.I. Nekrasov, \textit{Izv. Ivanovo-Vosnosonk.
Politehn Inst.} 6:155-71 (1922)\medskip

\noindent 4. A.I. Nekrasov,  The exact theory of steady state waves on
the  surface of a heavy liquid.
Technical Summary Report No 813. Mathematical Research center, University of
Wisconsin, 1967 [D.V. Thampuran, translator:C.W. Cryer, editor] \medskip

\noindent 5. L.M. Milne-Thompson, \textit{Theoretical Hydrodynamics},
Macmillan, London, 1968.\medskip

\noindent 6. G.A. Chandler and I.G. Graham,
The Computation of water waves modelled by Nekrasov's Equation.
\textit{SIAM J. Numer. Anal.} 30: 1041-1065 (1993).\medskip
\newpage 

\noindent \textbf{Figure Captions}

\noindent Table 1. Positions of the turning points, $s_t$  and the corresponding values,
$\phi (s_t)$ and comparison with those obtained by Chandler and Graham.

\noindent Table 2. The positions, $x_i$ of the zeros and the turning points of $s(\phi(s)-\pi/6) $
as a function of $x=\beta \log s $ and the corresponding values at the turning
points.
$\Delta x_i$ is the difference $x_i-x_{i-1}-\frac12$

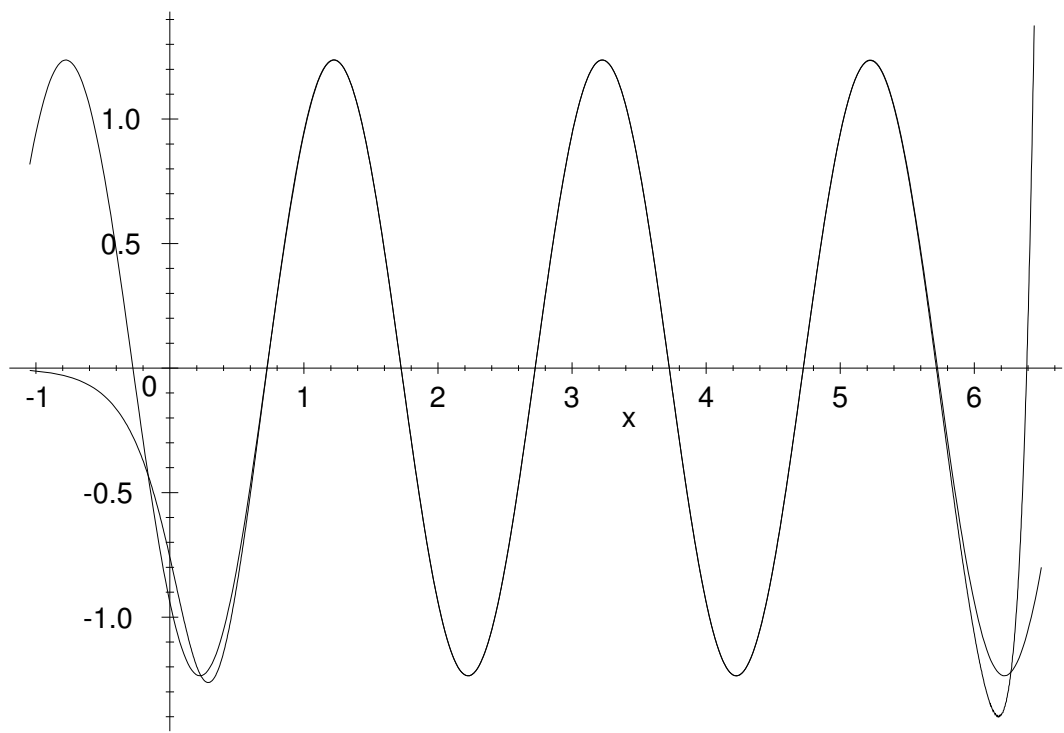
\noindent Figure 1. \rm Comparison $\Psi(x) \equiv (6\Theta(x) /\pi -1)s$ with
A$\sin(\pi(x -x_0))$ as a function of $x=\beta \log s/\pi$. 

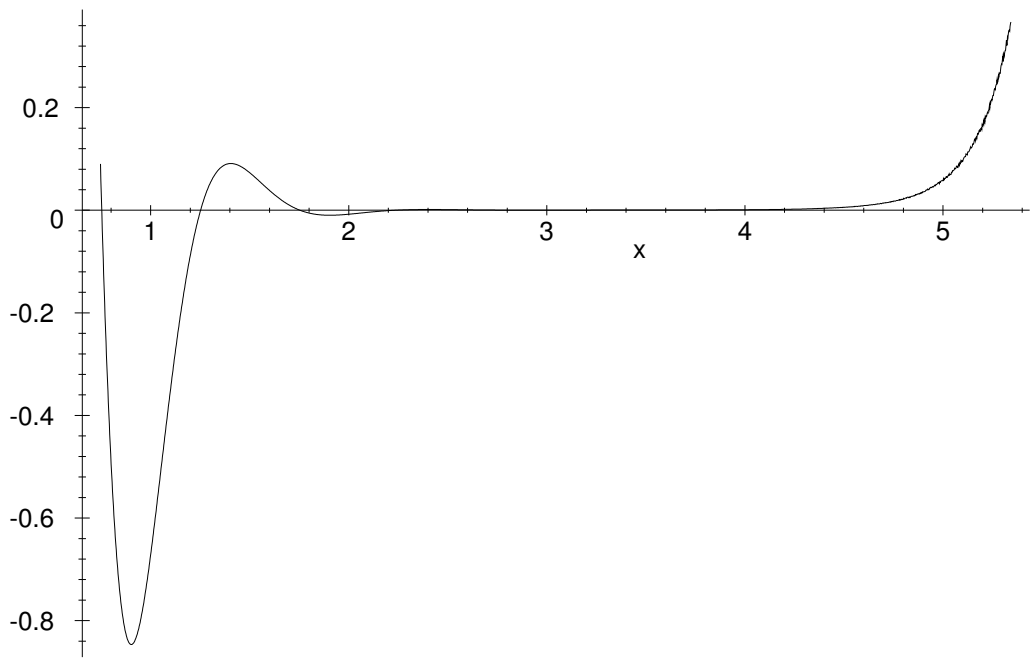
\noindent Figure 2. \rm Difference between the solution and its Asymptotic form
$100(\Psi(x) -A\sin(\pi(x -x_0)))$ as a function of $x=\beta \log s/\pi$. 

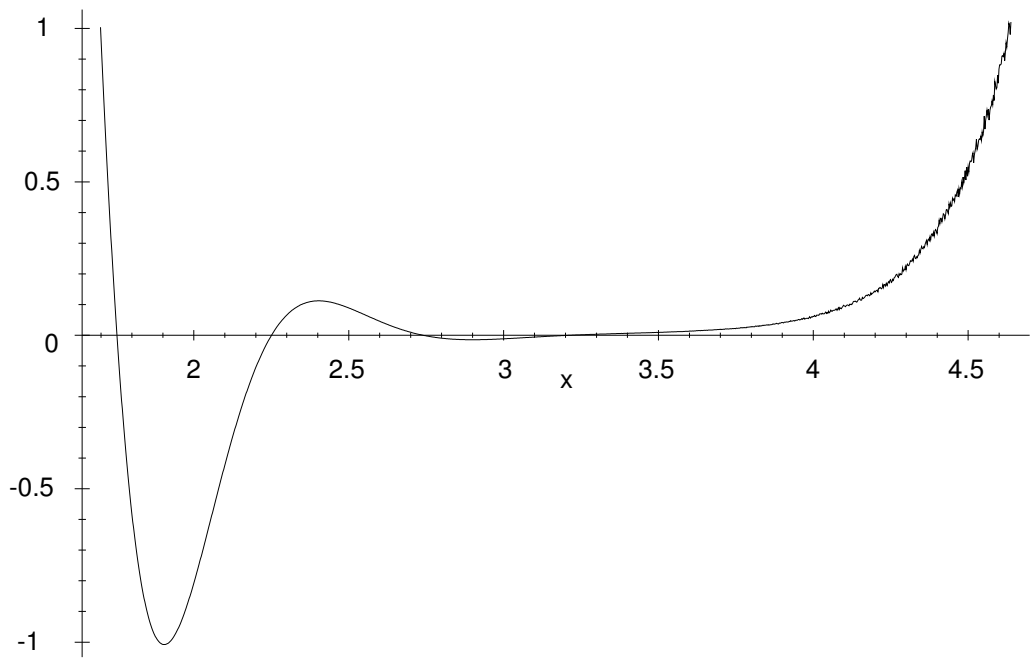
\noindent Figure 3. \rm Difference between the solution and its Asymptotic
form
$10000(\Psi(x) -A\sin(\pi(x -x_0)))$ as a function of $x=\beta \log s/\pi$. 

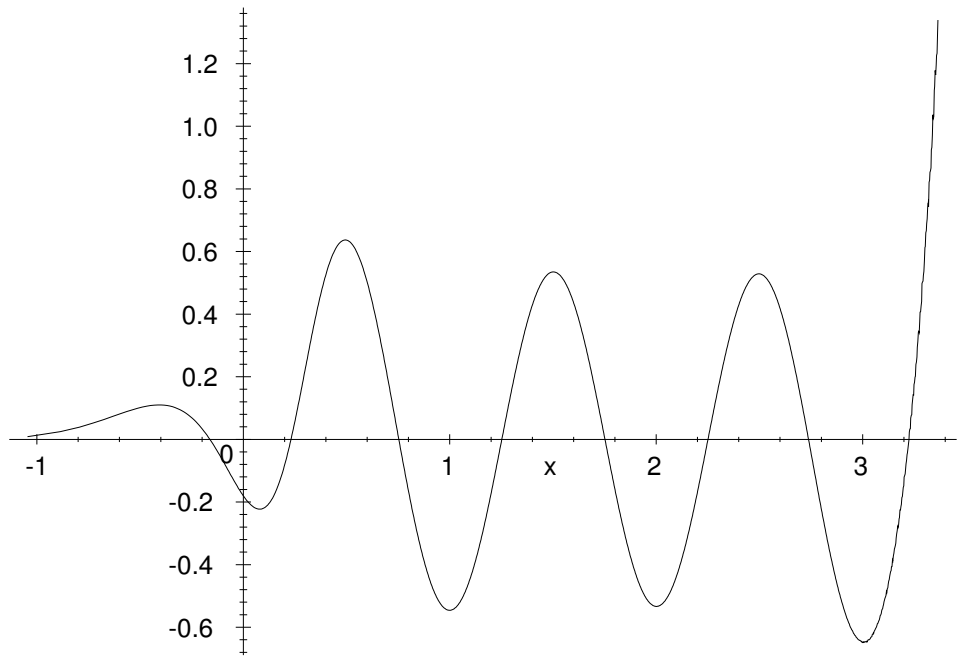
\noindent Figure 4. \rm Difference between the solution and its Asymptotic
form $\Psi_1(x) \equiv s((6\Theta(x) /\pi -1)s-A\sin(\pi(x -x_0)))$ 
as a function of $x=\beta \log s/\pi$. 

\end{document}